\documentclass[aps,twocolumn,preprintnumbers,groupedaddress,nofootinbib]{revtex4-1}
\usepackage{graphicx}
\usepackage{dcolumn}
\usepackage{bm,amsfonts,amsthm,amsmath,amssymb}
\usepackage[]{epsfig,graphics}
\usepackage{comment}
\usepackage[T1]{fontenc}
\usepackage{lipsum}
\usepackage[utf8]{inputenc}
\usepackage{ulem}
\usepackage{multirow}
\usepackage{color}
\usepackage{lastpage}
\usepackage[sort&compress]{natbib}
\usepackage{enumerate}
\usepackage{subfigure}
\usepackage{wasysym}
\usepackage{hyperref}
\usepackage{relsize}
\usepackage{physics}
\usepackage{mathrsfs}
\usepackage{tensor}

\hypersetup{
    bookmarks=true,         
    unicode=false,          
    pdftoolbar=true,        
    pdfmenubar=true,        
    pdffitwindow=false,     
    pdfstartview={FitH},    
    pdftitle={My title},    
    pdfauthor={Author},     
    pdfsubject={Subject},   
    pdfcreator={Creator},   
    pdfproducer={Producer}, 
    pdfkeywords={keyword1} {key2} {key3}, 
    pdfnewwindow=true,      
    colorlinks=false,       
    linkcolor=red,          
    citecolor=green,        
    filecolor=magenta,      
    urlcolor=cyan           
}

\newenvironment{itemize*}
  {\begin{itemize}
    \setlength{\itemsep}{0pt}
    \setlength{\parskip}{0pt}}
  {\end{itemize}}

\newenvironment{enumerate*}
  {\begin{enumerate}
    \setlength{\itemsep}{0pt}
    \setlength{\parskip}{0pt}}
  {\end{enumerate}}

\newenvironment{description*}
  {\begin{description}
    \setlength{\itemsep}{0pt}
    \setlength{\parskip}{0pt}}
  {\end{description}}

\def\ben{\begin{enumerate*}}
\def\een{\end{enumerate*}}
\def\bi{\begin{itemize*}}
\def\ei{\end{itemize*}}
\def\bd{\begin{description*}}
\def\ed{\end{description*}}
\def\be{\begin{equation}}
\def\ee{\end{equation}}
\def\bea{\begin{eqnarray}}
\def\eea{\end{eqnarray}}
\def\bfl{\begin{flushleft}}
\def\efl{\end{flushleft}}

\begin{document}

\title{A Preferred Mass Range for Primordial Black Hole Formation\\
and Black Holes as Dark Matter Revisited}

\author{Julian Georg}
\email{jsgeorg@syr.edu} 
\author{Scott Watson}
\email{gswatson@syr.edu} 
\affiliation{Department of Physics, Syracuse University, Syracuse, NY 13244, USA}

\date{\today}

\begin{abstract}
Bird, {\it et. al.} \cite{Bird:2016dcv} and Sasaki, {\it et. al.} \cite{Sasaki:2016jop} have recently proposed the intriguing possibility that the black holes detected by LIGO could be all or part of the cosmological dark matter.
This offers an alternative to WIMPs and axions, where dark matter could be comprised solely of Standard Model particles.
The mass range lies within an observationally viable window and the predicted merger rate can be tested by future LIGO observations.
In this paper, we argue that non-thermal histories favor production of black holes near this mass range -- with heavier ones unlikely to form in the early universe and 
lighter black holes being diluted through late-time entropy production.   We discuss how this prediction depends on the primordial power spectrum,
the likelihood of black hole formation, and the underlying model parameters. We find the prediction for the preferred mass range to be rather robust assuming a blue spectral index less than two.
We consider the resulting relic density in black holes, and using recent observational constraints, establish whether they could account for all of the dark matter today. 
\end{abstract}
\maketitle
\thispagestyle{empty}
\section{Introduction}
The question ``Did LIGO detect dark matter?'' was raised recently by the authors of \cite{Bird:2016dcv}. Following the first LIGO observation of gravitational waves GW150914 \cite{LIGO:2016}, the authors examined the possibility that all of dark matter could consist of 30 solar mass black holes. Making this assumption, they calculated the expected merger rate from binaries forming in dark matter halos. Instead, in \cite{Sasaki:2016jop} the authors examine a similar scenario (see also \cite{Orlofsky:2016vbd,Carr:2016drx}), but allow the fraction of dark matter that is comprised of primordial black holes (PBHs) to be a free parameter.  In both papers, it is noted that there are considerable astrophysical uncertainties associated with merger rates and the dark matter distribution, but given the uncertainties, their results can be consistent with the value measured by LIGO. Curiously, existing observational constraints leave a window in precisely this mass range (near a solar mass) to allow black holes to comprise all of the dark matter \cite{Bird:2016dcv}. However, one important theoretical piece to the story is lacking: {\it why are black holes in this mass range favored over other masses?}  We will argue that non-thermal histories offer a possible explanation.

A generic prediction of beyond the Standard Model (BSM) physics is the existence of additional scalar fields (moduli), which frequently predict an epoch of early matter domination following inflation. 
In many approaches the mass and decay rate of these scalars is determined by symmetry breaking processes near the TeV scale -- the physics of which has been well-studied, particularly for its implications for WIMPs and axions, collider physics, and fundamental questions such as the electroweak hierarchy and the Cosmological Moduli Problem (see \cite{Kane:2015jia} for a review and guide to the literature). 

In this paper, we are interested in another prediction of non-thermal histories during the post-inflationary epoch. Unlike in a thermal history, during a matter phase, PBHs can form on all sub-Hubble scales, resulting in a continuum of PBH masses \cite{Georg:2016yxa,Cotner:2016cvr,Khlopov:2008qy,PK81,Khlopov:1980mg,KP81,Harada:2016mhb}.
Depending on the primordial power spectrum, this can lead to strong constraints on these non-thermal epochs and the underlying particle theory \cite{Georg:2016yxa}.
Another aspect of these models is that the phase ends when the scalar decays, again `reheating' the universe. The associated entropy production dilutes any pre-existing relics (such as baryons or dark matter), but the decay can also be a replenishing source of baryons \cite{Allahverdi:2010rh,Kane:2011ih} or dark matter \cite{Moroi:1999zb,Acharya:2008bk}.  Here, we examine the effect of the non-thermal epoch and scalar decay on the formation and survival of PBHs as dark matter.

The rest of the paper is organized as follows. In the next section, we review the formation of PBHs in an early matter phase and the resulting prediction for the dark matter relic density taking into consideration entropy production. In Section \ref{prefer}, we establish an argument for a preferred mass range for the PBHs, and calculate the corresponding PBH abundance for each allowed mass. We then establish constraints on the allowed values of the spectral index of the primordial power spectrum, which is important for establishing both the mass and abundance of the most dominant PBHs today. We present our conclusions in Section \ref{conclude}. 

\section{Black Hole Formation in Early Matter Phases  }
What sets a non-thermal universe apart from a thermal universe is that PBHs can continue to form after perturbations cross the Hubble radius. Matter (scalar) perturbations will continue to grow because the Jeans pressure is negligible during epochs of scalar oscillations. This means that the usual radiation domination calculations for PBH formation are not applicable.

In matter domination, the probability of forming PBHs stems from two different considerations. Firstly, the over-dense region must collapse to its Schwarzschild radius before a caustic can form at its center (from which particles falling inwards might be dispersed and prevent the formation of a PBH).  This is also related to the need for the over-density to be sufficiently homogeneous once the perturbations break away from the background expansion (when they become non-linear) \cite{Khlopov:2008qy}.  This implies
\begin{equation}\label{eq:caus}
t_{\mbox{\tiny{PBH}}}\leq t_{\mbox{\tiny{caustic}}},
\end{equation}
where $t_{\mbox{\tiny{PBH}}}$ is the PBH formation time and $t_{\mbox{\tiny{caustic}}}$ is the time it takes a caustic to form.
In \cite{PK81} it is shown that the probability that \eqref{eq:caus} is obeyed can be approximated by 
\begin{equation}
W_{\mbox{\tiny{caustic}}}\simeq \delta_{\mbox{\tiny{M}}}^{3/2},
\end{equation}
where the density contrast is given by
\be
\label{eee1}
\delta_{\mbox{\tiny{M}}} \equiv \frac{\delta M}{M} \sim M^{\frac{(1-n)}{6}},
\ee 
where $n$ is the spectral index.

The second consideration takes into account if the gravitational collapse is sufficiently spherical and therefore results in a point like singularity and not a 1-dimensional (cigar) or 2-dimensional (Zel'dovich pancake) singularity. This probability was calculated in \cite{Doroshkevich1970} and is 
\begin{equation}\label{W_s}
W_s \simeq C \, \delta_{\mbox{\tiny{M}}}^p.
\end{equation}
This result comes from analytically evolving the initial perturbations to the non-linear regime and then tracking their behavior through modeling the collapse as a Tolman-like solution.
One can introduce a deformation tensor with random values (reflecting the nearly Gaussian initial spectrum) and then use this to find the probability of spherical collapse. 
After diagonalizing this tensor and finding the corresponding eigenvalues one arrives at the likelihood for PBH formation \cite{KP81}. Following \cite{Doroshkevich1970} the authors of \cite{KP81} find $p=5$ and $C=2\times 10^{-2}$. In what follows we will allow for a range of values for $C$ and $p$ and see that this uncertainty will not alter our main conclusions.

The mass fraction in PBHs of mass $M$ is then given by the total probability for their formation by  
\begin{equation} \label{bigbeta}
\beta(M)=\frac{\rho_{\mbox{\tiny PBH}}(M)}{\rho_{\mbox{\tiny tot}}} \simeq W_sW_{\mbox{\tiny{caustic}}} \simeq  C \, \delta_{\mbox{\tiny{M}}}^{p} \, \delta_{\mbox{\tiny{M}}}^{3/2}.
\end{equation}
In a radiation dominated universe $\rho_{\mbox{\tiny tot}}\sim 1/a^4$, whereas for PBHs which have not evaporated ($M>10^{16}$ g) we have $\rho_{\mbox{\tiny PBH}} \sim 1/a^3$.  Thus, the mass fraction grows as $\beta \sim a(t)$. However, in a matter dominated phase $\beta$ remains constant until the time of reheating
as long as there is not significant entropy production. Notably, in the scalar dominated phases considered here, entropy production plays an important role, which we now discuss.

\subsection{Entropy Production in a Non-thermal History}
In this section, we establish the effect of entropy production on PBH formation.
To be specific, we will focus on models where the non-thermal phase results from the displacement of scalar moduli following inflationary reheating (see \cite{Kane:2015jia} for a review).  Following reheating, the universe is radiation dominated. However, the scalar oscillations will quickly come to dominate the energy density since their energy density dilutes more slowly than radiation. The non-thermal phase begins around the time when the scalar oscillations become comparable to the radiation density and we will denote the radiation density at this time by $\rho_r^{(0)}$.  Then, the radiation density during the non-thermal epoch will be given by $\rho_r=\rho_r^{(0)}+\rho_{\mbox{\tiny gen}}$, where the additional contribution represents the radiation generated by scalar decays\footnote{We follow closely \cite{Gorbunov:2011zz} and refer to that textbook for a more detailed discussion.}.

The relevant equations are
\bea
\dot{\rho}_m+3H\rho_m&=&-\Gamma \rho_m, \label{ee1}\\
\dot{\rho}_r+3H\rho_r&=&\Gamma \rho_m, \label{ee2}\\
3H^2m_p^2&=&\rho_m+\rho_r \label{ee3}.
\eea
We can solve \eqref{ee1} 
\be
\rho_m = \rho_m^{(0)} \left( \frac{e^{-\Gamma t}}{a^3} \right).
\ee
We are interested in establishing when entropy production first becomes important and so we consider times $t_\ast < t \ll \Gamma^{-1}$.  Thus, we have $\Gamma t \ll 1$ 
and so $\rho_m \sim a^{-3}$ or $\rho_m = 4m_p^2 / (3t^2)$.  This corresponds to a matter dominated universe\footnote{This is not strictly correct since the radiation gives an equivalent contribution at $t_\ast$.  However, as is the case with the textbook treatment of standard radiation/matter equality it is a very good approximation and adequate for our purposes here (see e.g. Chapter 2 of \cite{padmanabhan1993structure}).} 
$H\simeq 2/(3t)$. Using this solution in \eqref{ee2} and solving, we find
\bea \label{rads}
\rho_r &=& \frac{4m_p^2 t_\ast^{2/3}}{3t^{8/3}}+\frac{4m_p^2 \Gamma}{5t}, \nonumber\\
&\equiv& \rho_{\mbox{\tiny init}}(t)+\rho_{\mbox{\tiny gen}}(t)
\eea
where $\rho_{\mbox{\tiny init}}$ and $\rho_{\mbox{\tiny gen}}$ are the initial radiation and radiation generated from scalar decays, respectively.  In arriving at \eqref{rads} 
we have used that  $\rho_m(t_\ast)=\rho_r(t_\ast)$ at the beginning of the non-thermal phase and 
$\rho_{\mbox{\tiny gen}}(t_\ast)=0$.

The production of entropy then becomes important when the amount of radiation generated is comparable 
to the initial radiation.  Setting $ \rho_{\mbox{\tiny init}}(t)=\rho_{\mbox{\tiny gen}}(t)$ we find this occurs at a time
$t=t_\ast^{2/5} \Gamma^{-3/5}$ and so entropy production is important long before $t=\Gamma^{-1}$.

Prior to this, any relics that have `frozen-out' (such as baryons or dark matter) will have constant co-moving number density
$Y\equiv n_x/s$ where $n_x$ is their number density and $s$ the entropy density.
However, once entropy production becomes important, 
any existing relics will be diluted 
until entropy production ceases near $\Gamma^{-1}$. 
The resulting abundance is then given by $Y_x \rightarrow Y_x \Delta^{-1}$. With $\Delta$ given by
\bea
\Delta &\equiv& \left. \frac{s_{\mbox{\tiny gen}}}{s_{\mbox{\tiny init}}} \right\vert_{t=\Gamma^{-1}}= \left. \left(\frac{\rho_{\mbox{\tiny gen}}}{\rho_{\mbox{\tiny init}}}\right)^{3/4} \right\vert_{t=\Gamma^{-1}}, \nonumber \\
&=&\left(\frac{3}{5}\right)^{3/4} \left( t_\ast \Gamma \right)^{-1/2}, \label{delta}
\eea
where we used $s \sim \rho_r^{3/4}$ and \eqref{rads}. If the co-moving number density instead became constant at a time $t_f>t_\ast$ then $t_\ast \rightarrow t_f$
in \eqref{delta}. 

\subsection{Entropy Production and the PBH Abundance \label{dilute}}
The previous section showed that one effect of entropy production is to dilute any existing relics.
The aim of this section is to establish the effect on the abundances of PBHs. 

We can account for the effect of entropy production on the mass fraction of PBHs using \eqref{delta}.
As discussed above, for the PBHs forming during the matter phase, the initial mass fraction at the time of PBH formation $\beta(t_f)$ will remain constant until the time of reheating -- except for the effect of entropy production.  Thus, using \eqref{delta}, for PBHs forming at $t_f$ the corresponding mass fraction at the time of reheating is given by
\bea
\beta(t_r) &=& \frac{\beta(t_f)} {\Delta}= \left( \frac{5}{3} \right)^{3/4} \left( \Gamma \, t_f \right)^{1/2}  \beta(t_f), \nonumber \\
&=& 3.8 \times 10^{-2} \left( \frac{g(T_r)}{10.75}\right)^{1/4}  \left( \frac{T_r}{5 \; \mbox{MeV}}\right)
\nonumber \\ && \times\left( \frac{M}{M_\odot}\right)^{1/2}  \left( \frac{t_f}{t_H}\right)^{1/2}   \beta(t_f), 
\label{theone}
\eea
where $T_r \sim g^{-1/4}(\Gamma m_p)^{1/2}$ is the reheat temperature and $t_f$ is the time of PBH formation with $\beta(t_f)$ the corresponding mass fraction given by \eqref{bigbeta}. We have also introduced the horizon crossing time $t_H$. It is important to emphasize that in the matter phase, PBHs can form after Hubble radius crossing and so, in general, for a given mode $t_f > t_H$. We have left this dependence in Eq.~(\ref{theone}) to account for this effect (in the previous literature it is commonly assumed $t_f = t_H$, but this is only an accurate approximation when $w=p/\rho \neq 0$). We have chosen fiducial values that are representative of low reheat scenarios and normalized the PBH mass to the solar mass $M_\odot = 2.0 \times 10^{33}$ g.
We will justify these choices shortly.

Following reheating due to the scalar decay, the co-moving density in PBHs, $Y_{\mbox{\tiny PBH}}$, will remain fixed. We can relate this to the mass fraction at reheating
\be
\beta(t_r) = \frac{4M}{3T_r} \left.\left( \frac{n}{s} \right)\right\vert_{t=t_r},
\ee
or 
\be \label{yield}
Y_{\mbox{\tiny PBH}}\equiv \frac{n}{s}=\left( \frac{3T_r}{4M} \right) \beta(t_r),
\ee
where we used $\rho \sim s T$ after reheating. We can then find the critical density in PBHs (for a given mass $M$) today
\be
\Omega_{\mbox{\tiny PBH}} = \frac{\rho_{\mbox{\tiny PBH}}}{\rho_c}=\frac{M Y_{\mbox{\tiny PBH}}s}{\rho_c},
\ee
where $\rho_c$ is the critical density. Utilizing \eqref{theone} and \eqref{yield} we then have
\bea \label{critdensity}
\Omega_{\mbox{\tiny PBH}} h^2&=& 0.1 \; \left( \frac{g(T_r)}{10.75}\right)^{1/4}  \left( \frac{T_r}{5 \; \mbox{MeV}}\right)^2 \left( \frac{M}{M_\odot}\right)^{1/2} \nonumber \\
&\times&   \left( \frac{t_f}{t_H}\right)^{1/2}   \left(\frac{\beta(t_f)}{2.5 \times 10^{-6}}\right), 
\eea
where $\beta(t_f)$ is the undiluted mass fraction given by \eqref{bigbeta}, and $\rho_c/s_0=3.6 \times 10^{-9} h^2$ GeV, with $h$ the Hubble parameter in units of $100$ km$/$s$/$Mpc (we take $h=0.7$).

\subsection{Expected Mass Range}
Having established the effect of entropy production on the mass fraction and abundance of PBHs, we now consider the possible ranges of masses.  The minimum PBH mass is determined by the size of the Hubble volume at the beginning of the matter dominated phase.  Matter phases resulting from the oscillations of a scalar field \cite{Georg:2016yxa} start with the onset of oscillations when $H_{\mbox{\tiny osc}}\simeq m_{\sigma}$. The energy density at that moment within the Hubble volume is $\rho =M_{\mathrm{min}}H_{\mbox{\tiny{osc}}}^3$, and we have
\begin{equation}
M_{\mbox{\tiny{min}}}=\frac{3H_{\mbox{\tiny{osc}}}^2m_p}{H_{\mbox{\tiny osc}}^3}=3\frac{m_p}{H_{\mbox{\tiny{osc}}}}\simeq 3\frac{m_p}{m_{\sigma}} \label{min}.
\end{equation}
The maximum PBH mass is given by \cite{Georg:2016yxa}
\begin{equation}
M_{\mbox{\tiny max}} = \alpha^{\frac{1}{n+3}}  \left(\frac{M_{\mbox{\tiny C}}}{m_p}\right)^{\frac{n-1}{n+3}}\left(\frac{m_{p}}{m_{\sigma}}\right)^{\frac{12}{n+3}} \, m_{p},  
\label{max} 
\end{equation}
where $\alpha=3.6 \times 10^{-22}$, $M_{\mbox{\tiny C}}=10^{57} h^{-1} $ g ($1$ GeV = $1.8 \times 10^{-24}$ g). As explained in \cite{Georg:2016yxa}, this result comes from looking at the last PBHs to form at the time of scalar decay taking into account the evolution of perturbations on sub-Hubble scales as well. 
This is because the later in the universe a PBH is produced, the more massive it typically is, due to the increased size of the Hubble volume and the prolonged growth of the perturbations. The decay width of the scalar (moduli) is typically given by $\Gamma  = c \, m_{\sigma}^3 / m_p^2$, where $c$ is slightly model-dependent, but typically an order one number. The lighter the scalar is, the later it decays and the larger the maximum PBH mass. Therefore, we establish that the scalar mass driving the non-thermal phase determines both the minimal and maximal mass and thus the entire range of expected PBH masses. 

\section{Non-thermal Histories and a Preferred Mass Range for PBH Dark Matter \label{prefer}}
We have seen that the scalar mass $m_\sigma$ controls the range of PBH masses and also the reheat temperature $T_r \sim (\Gamma m_p)^{1/2} \sim m_\sigma^{3/2} /m_p^{1/2}$.
Therefore, the mass fraction \eqref{theone} and the PBH abundance \eqref{critdensity} basically depend on two {\it a priori} free parameters -- the mass of the scalar $m_\sigma$ and the spectral index $n$.
In this section, we argue that the mass is in fact not a free parameter in fundamental approaches to BSM physics that predict non-thermal histories.  
We then use the preferred mass to find the expected relic abundance of PBHs and use existing observations to place restrictions on the spectral index. 

\subsection{The PBH / Dark Matter Relic Density}
We have seen that PBH formation in a matter phase is vastly different than that in a thermal universe. We instead expect PBHs to  form over a range of masses $M_{\mathrm{min}}\lesssim M \lesssim M_{\mathrm{max}}$ determined by the scalar mass $m_\sigma$. In fundamental approaches to BSM, this mass is typically set by a scale of symmetry breaking (which must be at or above the Electro-weak scale) and, as such, is connected to other aspects of the BSM theory.  As a well-studied example, we briefly consider supersymmetry (SUSY), but  emphasize this expectation occurs in other approaches to BSM physics. \\
If SUSY is broken by gravity or anomaly mediation, the mass of the scalar is tied to the gravitino mass and the scale of SUSY breaking, as $m_\sigma \sim m_{3/2} \sim \Lambda^2 / m_p$,  where $\Lambda$ is the scale of SUSY breaking.   This same scale controls other aspects of the theory, such as the squark masses and the level of flavor changing neutral currents (FCNC).  Thus, the scalar mass is really a consequence of SUSY breaking, and so is the duration of the non-thermal phase \cite{Kane:2015jia}.  Observationally, existing bounds from the Large Hadron Collider (LHC) push squark masses to be considerably higher than the electro-weak scale, implying that the scalar (moduli) masses would also have to be higher for consistency with collider data.  However, such a `split-spectrum' (with TeV-scale masses) \cite{Wells:2004di,ArkaniHamed:2004yi,Arvanitaki:2012ps} was actually first proposed by Wells in 2004  \cite{Wells:2004di}. This was long before LHC constraints on SUSY, as it could improve SUSY's phenomenological successes (e.g. improved gauge coupling unification), while also addressing challenging issues for SUSY like FCNC and CP violation.  What was less appreciated at the time is that pushing the symmetry breaking scale slightly higher could also resolve the so-called Cosmological Moduli Problem \cite{Kane:2015jia}.  Allowing for this modest hierarchy in scales, raising the SUSY breaking to the TeV range then predicts scalar (moduli) masses around $m_\sigma \simeq 10 - 100$ TeV in a large range of BSM models \cite{Kane:2015jia}.  Although SUSY represents the best studied example, there are also string theory and extra-dimensional approaches to the hierarchy problem that make similar predictions for TeV-scale moduli. Independently, one could also simply ask the question, {\it what was the lowest possible reheat temperature in models of inflation?} \cite{Giudice:2000ex}. The answer to this question leads to a similar conclusion as the BSM considerations above. The universe must be thermalized by the time of BBN ($T_{\mathrm{r}} \simeq 5$ MeV), which corresponds to $m_{\sigma}\simeq 50$ TeV. 
\begin{figure}[t!] 
\begin{center}
\includegraphics[scale=0.35]{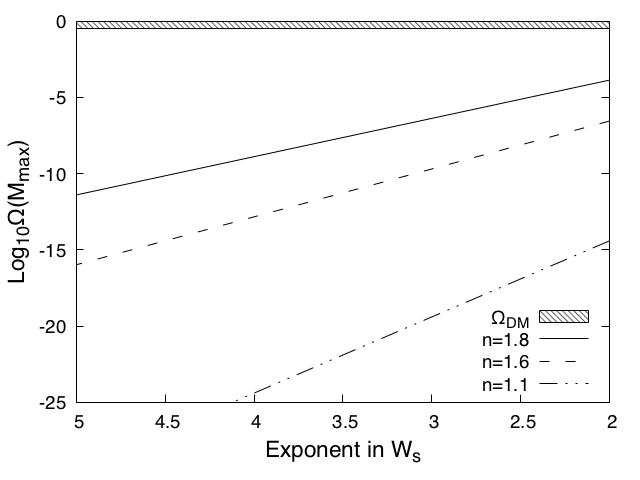}
\end{center}
\caption{\label{omega}We calculate the density parameter $\Omega_{\mbox{\tiny PBH}}$, \eqref{critdensity}, for PBHs of mass $M_{\mbox{\tiny max}}$ as a function of the exponent in the probability for spherical collapse, \eqref{W_s}, for different values of the spectral index $n$. The ceiling shown represents the current measured value for $\Omega_{\mbox{\tiny DM}}$. 
As discussed in the text, we allow for freedom in the exponent of the probability for spherical collapse $W_s$ (see \eqref{W_s}), which does make a significant difference in the abundance, but even allowing for this freedom we see the interesting case of $n=1.6$ will not lead to PBHs as the dominant source of dark matter.}
\end{figure}
\begin{figure*}
\begin{minipage}[t]{.45\textwidth}
\includegraphics[scale=0.33]{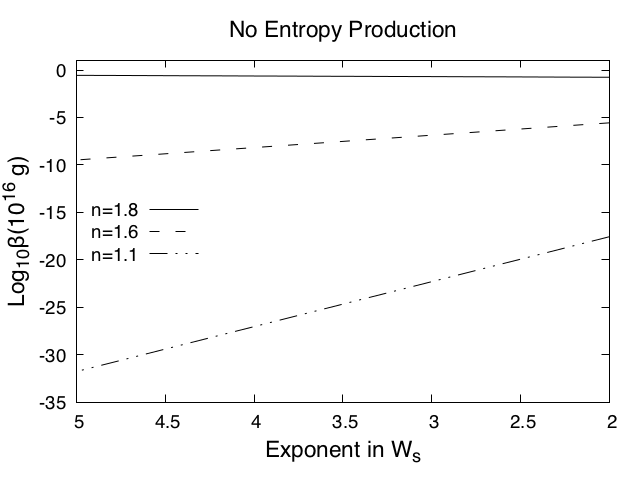}
\caption{The mass fraction $\beta$ of PBHs with mass $10^{16}$ g neglecting the entropy production associated with the decay of the scalar (modulus), for different values of the spectral index $n$ and reflecting the theoretical uncertainties in the formation probability $W_s$. Without dilution, a blue spectrum will generally overproduce these PBHs and be inconsistent with existing observations (compare with Fig.~\ref{beta_fermi}). }\label{beta_tf_fermi} 
\end{minipage}\qquad
\begin{minipage}[t]{.45\textwidth}
\includegraphics[scale=0.33]{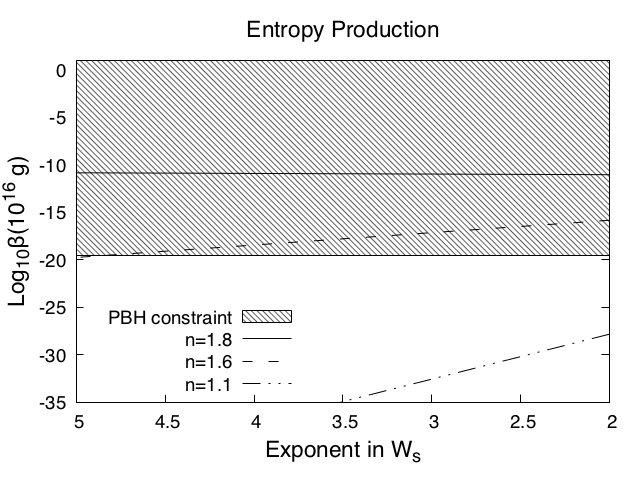}
\caption{The mass fraction $\beta$ for the same values as in Fig.~\ref{beta_tf_fermi}, but now accounting for dilution associated with the entropy production of the decaying scalar (see Eq.~(\ref{theone})).
Given the theoretical uncertainties, we see that a spectral index $n\gtrsim1.6$ is in tension with the data, whereas lower values avoid constraints due to the dilution.}\label{beta_fermi}
\end{minipage}
\end{figure*}

Given this motivation, we consider a $m_\sigma \simeq 50$ TeV scalar and find the corresponding prediction for the range of possible PBH masses.
The lightest PBHs are given by \eqref{min} and correspond to $M_{\mbox{\tiny min}} \simeq 10^9$ g.  
 The maximal mass is given by \eqref{max}. The exact value depends on the spectral index of the primordial power spectrum, which is the only remaining free parameter (although it is observationally restricted as we discuss shortly).  For a slightly blue spectral index ($1 < n<2$), we find a maximal PBH mass near the solar mass range, or slightly above.
This implies that in an early matter phase driven by scalar(s), prior to BBN, that 
PBHs will form in the mass range $10^9 \; \mbox{g} \lesssim M \lesssim 1000 M_\odot$. 

Our main finding is that, given this mass range, entropy production 
will dilute the lighter PBHs, since they form earlier, whereas the most massive PBHs form last, near the end of reheating. 
That is, from \eqref{theone} we see the amount of entropy dilution depends on the formation time as compared to the reheat time ($t_f/\Gamma^{-1})$.  This suggests that the last PBHs to form will be the most abundant.  Which corresponds to the maximal allowed PBH mass given by \eqref{max}. 

It is important to note that this maximal mass does depend on the spectral index. 
We find that for a blue spectrum (at pre-BBN scales) with $n=1.6$ that the maximal 
mass is then $24 \; M_\odot$ -- near the mass detected by LIGO and suggested by Bird, {\it et. al.} in \cite{Bird:2016dcv}. 
However, for lower (higher) values the maximal mass will also be lower (higher). 
The corresponding abundance today is then given by \eqref{critdensity}, also using \eqref{bigbeta}. 

Our results are presented in Fig. 1, where we consider  
a range of values for the spectral index. We also allow for uncertainties in the initial formation probability. As mentioned above,
this probability relies on estimating the evolution of perturbations in the non-linear regime, a task that is familiar from studies of large scale structure and is
typically done through N-body simulations. We are not aware of such studies prior to BBN (as a thermal history is typically assumed), and instead we rely on analytic estimates and allow for a range of 
uncertainty, where the classic literature \cite{Khlopov:2008qy,PK81,Khlopov:1980mg,KP81} found $\beta(t_f)\sim \delta_M^{13/2}$, 
and we instead find $\beta(t_f)\sim \delta_M^{7/2}$. Regardless of this uncertainty, we see from Fig. 1 that for $n=1.6$, 
these PBHs will be a small fraction of the dark matter. Most optimistically ($p=2, C=10^{-1}$ in \eqref{W_s}), we find $\Omega_{\mbox{\tiny PBH}}=2.8 \times 10^{-7}$.
The PBH contribution to dark matter increases with the spectral index, but as we discuss next, this also implies stronger tension with observations due to the abundance of lighter PBHs.

\subsection{Constraints on the Spectral Index}
The spectral index on the scales we are considering is unknown. In this sense, we have traded the question of what PBH mass is preferred for the unknown value of the spectral index.
However, we find that although entropy dilution can significantly reduce the number of PBHs with $M<M_{\mbox{\tiny max}}$, increasing the spectral index will increase the 
formation probability of lighter PBHs as seen from \eqref{eee1} and \eqref{bigbeta}. Thus, we can use existing observations to place an upper bound
on the spectral index.  In Figs. \ref{beta_tf_fermi} and \ref{beta_fermi} we focus on $M_{\mbox{\tiny PBH}}=10^{16}$ g PBHs, since these are within the mass range $10^{15}$ g -- $10^{17}$ g most strongly restricted by CMB and $\gamma$-ray observations \cite{Carr:2009jm}.  In Fig. \ref{beta_tf_fermi} we present the formation probability for differing values of the spectral index reflecting the theoretical uncertainties. We present the same predictions in Fig. \ref{beta_fermi}, but taking into account the effect of entropy dilution and we include the most recent observational constraints \cite{Clark:2016nst}. 
We find that even with entropy dilution and the uncertainties in the formation probability, that $n>1.6$ would be in conflict with existing observations.  Instead we find that $n=1.6$ is marginally compatible with existing data given the theoretical uncertainties in the formation process. It is somewhat intriguing that this corresponds exactly to the maximal mass range corresponding to $M_{\mbox{\tiny PBH}}\simeq 30 \; M_{\odot}$.  However, we realize this is both speculative, and we also emphasize again that even in this case we find that PBHs would not make up all the cosmological dark matter. 

\section{Conclusions \label{conclude}}
In this paper we have considered the process of PBH formation in a non-thermal history. Assuming the non-thermal phase results from 
the oscillations of a $10-100$ TeV scalar field (which is well-motivated by BSM physics) 
we have a prediction that PBHs will form in a mass range $10^9 \; \mbox{g} \lesssim M \lesssim 1000 M_\odot$, 
although the maximal mass depends on the spectral index of the primordial power spectrum.
If we then account for the effect of entropy production as the scalar decays, we find the most abundant PBHs will be the heaviest.
Other PBHs will form, and this leads to a restriction on the spectral index of around $n<1.6$.
We find that for a spectral index near $n=1.6$ that non-thermal histories predict that the most abundant 
PBHs will be around $M_{\mbox{\tiny PBH}} \simeq 30 \; M_{\odot}$ -- motivating the proposal of \cite{Bird:2016dcv} -- however, we also find they would only be a small fraction of the total amount of dark matter.
Even with such a small relic density, it should still be possible to further establish model building constraints using Pulsar timing as discussed recently in \cite{Orlofsky:2016vbd}.
We leave this investigation to future work.
\\

\section*{Acknowledgements}
We thank Simeon Bird, Steven Clark, Bhaskar Dutta, Yu Gao, Nicholas Orlofsky, and Louis Strigari for useful discussions.  
S.W. thanks the Michigan Center for Theoretical Physics for hospitality.
This work was supported in part by NASA Astrophysics Theory Grant NNH12ZDA001N and DOE grant DE-FG02-85ER40237.

\bibliographystyle{apsrev4-1}

\end{document}